
\vsize=7.5in
\hsize=6.6in
\hfuzz=20pt
\tolerance 10000

\baselineskip 12pt plus 1pt minus 1pt
\pageno=0

\def\){]}
\def\({[}

\def\rjustline#1{\line{\hss#1}}

\rjustline{WIS-94/24/May-PH}
\rjustline{TAUP 2165-94}

\centerline{\bf How to Use
 Weak Decays in Analyses of Data on Nucleon Spin Structure Functions}

\author{Harry J. Lipkin}
\smallskip
\centerline{Department of Nuclear Physics}
\centerline{\it Weizmann Institute of Science}
\centerline{Rehovot 76100, Israel}

\centerline{and}

\centerline{School of Physics and Astronomy}
\centerline{Raymond and Beverly Sackler Faculty of Exact Sciences
}
\centerline{\it Tel Aviv University}
\centerline{Tel Aviv, Israel}
\centerline{May 19, 1994}
\abstract

The use of weak decays to determine proton spin structure is examined in
view of possible violations of the Bjorken and Gottfried Sum rules,
flavor symmetry breaking and flavor asymmetry in the sea. The use of the
neutron decay is found to be unaffected by all these. A method for
including these effects in analyses of hyperon decays shows that a
flavor-asymmetric sea produced by SU(3) symmetry breaking has only a
small effect on results for the total spin carried by quarks. However
the strange quark contribution cannot be reliably obtained from charged
lepton scattering and weak decay data alone, and requires additional
model-dependent input relating nucleon and hyperon wave functions.
\endpage

Recent analyses of data on nucleon spin structure
\REF{\SMC}{The Spin Muon Collaboration (SMC),
CERN preprint CERN-PPE/93-206}
\REF{\ELKAR}{John Ellis and Marek Karliner,
Phys. Lett. B313 (1993) 131, and CERN preprint CERN-TH.7022/93,
Tel Aviv preprint
TAUP 2094-93 hep-ph/9310272, to be published in the Proceedings of
PANIC 93, Perugia, Italy}
$[{\SMC},{\ELKAR}]$
confirm the originally surprising conclusion
\REF{\BEK}{Stanley J. Brodsky, John Ellis and Marek Karliner, Phys. Lett.
B206 (1988) 309}
$[{\BEK}]$ of zero quark spin contribution to the proton spin but
continue to use semileptonic weak decay data to determine the fractional
contributions of the $u$, $d$ and $s$ - flavored current
quarks and antiquarks respectively to the spin of the proton
conventionally denote by $\Delta u (p) $, $\Delta d (p) $ and
$\Delta s (p) $.
However, the weak decays measure flavor-changing transition matrix
elements which have been noted\REF{\LIP}{Harry J. Lipkin, Phys. Lett.
B230 (1989) 135} $[{\LIP}]$ to
depend upon the wave functions of both the initial and final states. They
therefore give information about both nucleon and hyperon spin structures
which cannot be disentangled without some hyperon model.
Serious difficulties and paradoxes arise in all attempts fit the
observed hyperon magnetic moments and weak decays with models for hyperon
spin structure\REF{\PNAGOYA}{Harry J. Lipkin,
in ``Frontiers of High Energy Spin Physics
Proceedings of the 10th International Symposium on High Energy Spin
Physics, November 9-14 1992, Nagoya, Japan"
Edited by T. Hasegawa N. Horikawa, A Masaike and S. Sawada,
Universal Academic Press (1993) p. 701}
$[{\PNAGOYA}]$. In particular
the values for the ratio $\Delta s (\Sigma) /\Delta s (\Lambda) $
determined by fitting magnetic moment and semileptonic decay data
are not simply inconsistent. They
differ by a factor of $8 \pm 2$ and it is very difficult to find any
simple correction which can fix such a large factor.
Until these difficulties are resolved all values quoted for
$\Delta s (p) $ based on using hyperon data with assumptions about
$\Delta s (\Sigma) $ and $\Delta s (\Lambda) $ should be viewed with
suspicion. With this caveat, we attempt to do a bit better than the
conventional treatments
$[{\ELKAR},{\BEK}]$
which assume
very simple relations between nucleon and hyperon wave functions; e.g.
that the $\Sigma^-$ wave function is the mirror of the neutron wave
function under $u \leftrightarrow s $ flavor exchange. The possibility
of measuring $\Delta u (p)- \Delta d(p) - \Delta s(p)$ directly
by elastic neutrino scattering is now under consideration$[{\ELKAR}]$.
We do not consider this further here.

More general relations between weak decays and baryon spin structure
$[{\LIP}]$ were obtained from the current
algebra of the electroweak standard model. The initial and final states
are assumed to dominate the sum over intermediate states in the
evaluation of current commutators, but mirror symmetry is not assumed.
$$ {{G_A}\over{G_V}}(n\rightarrow p) \approx
{{\bra p \Delta u - \Delta d \ket p - \bra n \Delta u - \Delta d \ket n}
\over{2}}
\eqno (1a) $$
$$ {{G_A}\over{G_V}}(\Sigma^-\rightarrow n) \approx
{{\bra n \Delta u - \Delta s \ket n -
\bra {\Sigma^-} \Delta u - \Delta s \ket {\Sigma^-}}
\over{2}}
\eqno (1b) $$
These results are exact in the mirror symmetry limit where
the two terms on the right hand side are equal and give
information about nucleon spin structure.

The isospin symmetry relevant for neutron decay is
valid to the approximation needed for proton
spin structure, and completely immune to effects producing violations
of the Bjorken and Gottfried sum rules. The Bjorken sum rule relates
deep inelastic scattering data to the neutron decay and depends upon QCD
to obtain the relevant information from the deep inelastic data. Its
failure would reflect on QCD, not on the relations between neutron decay
and proton spin structure. The Gottfried sum rule assumes an isospin
symmetric sea. Its failure would indicate a charge asymmetry in the sea
produced presumably by charge exchange interactions between valence and
sea quarks; e.g. by pion or $\rho$ exchange, and in any case by strong
interactions which conserve isospin. Thus the neutron and proton remain
isospin mirrors under $ u \leftrightarrow d$ even with flavor asymmetric
seas. Substituting this mirror symmetry into eq.(1a) gives:

$$ N\equiv {{G_A}\over{G_V}}(n\rightarrow p) = F + D = $$
$$ = \Delta u (p) - \Delta d (p) =
\Delta d (n) - \Delta u (n)
= 1.2573 \pm 0.0028
\eqno (2) $$
where we have denoted this quantity by $N$ to simplify further equations,
introduced the $F$ and $D$ parameters used in conventional treatments
$[{\ELKAR},{\BEK}]$
and substituted the experimental results\REF{\PDG}{Particle
Data Group, Phys. Rev. D45 (1992) S1}
$[{\PDG}]$.

The flavor-SU(3) symmetry relevant for $\Sigma^-\rightarrow n$ decay is
broken. We now correct the results of ref. $[{\ELKAR}]$ for the most
obvious symmetry breaking mechanism, the higher mass of the strange quark
which reduces the production of strange quark pairs by gluons in the sea
and therefore destroys the mirror symmetry under $u \leftrightarrow s$.
We assume that the sea contribution to $\Delta s$ is equally suppressed
in both states and therefore that
$\Delta u - \Delta s$ is the $same$ in both hadron seas, rather than
being mirror images with opposite signs, and that flavor polarization
effects analogous to those in the $u-d$ sector can be neglected.
A symmetry-conserving strangeness exchange between valence and sea; e.g.
$K$ or $K^*$ exchange would give a hyperon wave
function with nonstrange valence quarks and a sea containing an extra
strange quark in addition to $s \bar s$ pairs. Both strangeness exchange
and the introduction of an extra strange quark in the sea are expected
to be suppressed in comparison with the $\pi$ or $\rho$ exchange and the
extra $u$ or $d$ quark in the sea which could arise in the nonstrange
sector and give rise to violation of the Gottfried sum rule.

A flavor-asymmetric sea which is the $same$ for the neutron and
$\Sigma^-$ can
be treated\REF{\Sivers}{G. Ramsey et al Phys. Rev. D39 (1989) 361}
$[{\Sivers}]$ by decomposing the quantities
$\Delta u (p) $,
$\Delta d (p) $ and
$\Delta s (p) $
into valence and sea contributions
denoted by the superscripts $v$ and $s$ respectively.
Since the neutron contains no valence strange quarks, we
can write
$$ \Delta s^v (n) = 0; \;\;\;\;\;
\Delta u^s(\Sigma^-) = \Delta u^s(n) = (1+\epsilon) \Delta s(n)
\eqno(3) $$
where $\epsilon$ describes the flavor asymmetry of the sea in the
$u-s$ sector and has the same value for the neutron and the $\Sigma^-$.
Substituting this result into eq. (1b) gives
$$ \Sigma \equiv
{{G_A}\over{G_V}}(\Sigma^-\rightarrow n) =
\Delta u^v (n) - \Delta s^v (n) = \Delta s^v (\Sigma^-) -
\Delta u^v (\Sigma^-)
\eqno (4) $$
where we have used the shortened notation $\Sigma$.
Substituting eqs. (3) and the experimental
value $[{\PDG}]$ into (15) gives
$$   \Sigma = F - D = \Delta u (n) - \Delta s (n) -
\{\Delta u^s(n) - \Delta s(n)\} = $$
$$ = \Delta u (n) - (1+\epsilon) \Delta s (n) =
\Delta d (p) - (1+\epsilon) \Delta s (p)
= -0.340 \pm 0.017.
\eqno (5) $$

We can compare our treatment with the conventional SU(3) analyses
$[{\ELKAR},{\BEK}]$ of weak decays by noting that Eqs. (2) and (5)
give
$ F = 0.459 \pm 0.009$,
$ D = 0.799 \pm 0.009$ and
$$ {{2 \Sigma + N} \over{\sqrt 3}} =
{{3F - D} \over{\sqrt 3}} =
$$ $$ =
{{[ \Delta u (p)+ \Delta d (p) - 2(1+\epsilon)\cdot \Delta s (p)]}
\over{\sqrt 3}} =
0.356 \pm 0.020
\eqno (6a) $$
$$ F/D = 0.574 \pm 0.013
\eqno (6b) $$

These numerical results are essentially indistinguishable from
those obtained in the standard treatment $[{\ELKAR}]$; namely,
$0.34 \pm 0.020$ for the expression (6a) which transforms in the
SU(3) symmetry limit like the isoscalar component of an octet
and $ F/D = 0.58 \pm 0.02 $. This seems strange,
since we have only considered the $\Sigma^-\rightarrow n$ decay
and not used the additional input available from the experimental
data for the other two weak decays. We show below that errors on the
other decays are too large to have any impact on the $F$ and $D$ values
determined by fitting all the data.

Thus the effect of SU(3) breaking is simply incorporated
in the standard SU(3) analysis which determines the F and D parameters
from the $n\rightarrow p$ and $\Sigma^-\rightarrow n$ decays and uses
them to determine two independent linear combinations of $\Delta u$,
$\Delta d$ and $\Delta s$
which transform like the isovector and isoscalar components of an
SU(3) octet; namely (2) and (6a). One simply
replaces $\Delta s(p) $ by $(1+\epsilon) \Delta s(p)$.

We now introduce this SU(3)-breaking correction into
the standard analysis of the EMC experiment and weak decays which
begins with the relation
$[{\BEK}]$
$$
{4\over 9}\Delta u (p)+ {1\over 9}\Delta d (p)+ {1\over 9}\Delta s (p)=
E \eqno (7)$$
where E denotes an experimental number obtained from polarized deep
inelastic scattering data with various $QCD$ corrections $[{\BEK}]$
to the naive parton model result.

Solving the three equations (7), (2) and (5)
for $ \Delta u (p)$, $\Delta d (p)$ and $\Delta s (p) $  gives
$$ \Delta u = {1\over {6+5\epsilon}}\cdot \{9(1+\epsilon)E+(2+\epsilon)N
+ \Sigma\} \eqno (8a) $$
$$
\Delta d = {1\over {6+5\epsilon}}\cdot \{(1+\epsilon)(9E-4N) + \Sigma
\} \eqno (8b) $$
$$ \Delta s
= {1\over {6+5\epsilon}}\cdot \{9E-4N - 5\Sigma\}
= {6\over {6+5\epsilon}}\cdot \Delta s (\epsilon = 0)
\eqno (8c) $$
where we have suppressed the argument $(p)$
for all cases referring to the proton spin structure.
The total quark spin contribution to the proton spin is then
$$ \Delta q \equiv \Delta u + \Delta d + \Delta s =
{{9E-2N - \Sigma}\over 2}
\cdot \left(1 - {{\epsilon}\over {6+5\epsilon}} \right)
+\epsilon \cdot \left({{N + 2 \Sigma}\over {6+5\epsilon}}\right)
$$ $$ \approx
\Delta q (\epsilon = 0)
\cdot \left(1 - {{\epsilon}\over {6+5\epsilon}} \right)
+{{7\epsilon}\over {12(6+5\epsilon)}}
\eqno (9a) $$
where we have inserted the approximate experimental values
$N \approx (5/4)$ and $\Sigma \approx -(1/3)$
to give an estimate of the effects of SU(3) breaking and
$$ \Delta q (\epsilon = 0) \equiv {{9E-2N - \Sigma}\over 2}
\eqno (9b) $$
is the total contribution of quarks to the proton spin in the SU(3)
symmetry limit. The total contribution of the sea quarks is given by
$$ \Delta q^s \equiv
(3+2\epsilon)\cdot \Delta s
=3 \Delta s (\epsilon = 0)
\cdot \left(1 - {{\epsilon}\over {6+5\epsilon}} \right)
\eqno (9c) $$
The recent analysis $[{\ELKAR}]$ assuming SU(3) symmetry gives
$$ \Delta q (\epsilon = 0) = 0.12 \pm 0.17 \eqno (10a)   $$
$$ \Delta s (\epsilon = 0) = -0.19 \pm 0.06 \eqno (10b)  $$
from which we can extract the sea and valence contributions as
$$ \Delta q^s (\epsilon = 0) = -0.57\pm 0.18 \eqno (10c)   $$
$$ \Delta q^v (\epsilon = 0) = 0.69 \pm 0.25 \eqno (10d)   $$
Substituting $\epsilon=1$ into eqs. (8-10) gives the result for the
case of an asymmetric sea with $\Delta s$
reduced by a factor of 2 with respect to $\Delta d$,
$$
\Delta q (\epsilon = 1) =
\Delta u + \Delta d + \Delta s \approx 0.16 \pm 0.15
\eqno (11a) $$
$$
\Delta s (\epsilon = 1)
= (6/11)\cdot \Delta s (\epsilon = 0) = -0.10 \pm 0.03
\eqno (11b) $$
$$ \Delta q^s (\epsilon = 1) = -0.50\pm 0.15 \eqno (11c)   $$
$$ \Delta q^v (\epsilon = 1) = 0.66 \pm 0.21 \eqno (11d)   $$
The correction to $\Delta q$ is very small but $\Delta s$ is
significantly reduced.

The valence contributions are seen to be consistent with the
value (3/4) given by naive constituent quark models with parameters
adjusted to fit $ {{G_A}\over{G_V}}(n\rightarrow p) = 1.25 $. The sea
contributions nearly cancel the valence contributions to give a low
value for $\Delta q$. These results are very insensitive
to symmetry-breaking corrections.
However, the flavor composition of the sea is seen
to be not well determined by the data, and the value of
$\Delta s$ is sensitive to model-dependent SU(3) breaking corrections.

These qualitative features are illustrated by
a useful general inequality independent of the details of
SU(3) symmetry breaking. We
assume only that the value of $ \Delta s (\epsilon = 0) $
determined in the symmetry limit has the correct sign and that symmetry
breaking reduces its magnitude.
$$  0 \geq \Delta s \geq \Delta s (\epsilon = 0) \eqno (12) $$
It is then convenient to rewrite the expression (7)
\REF{\HJL}{Harry J. Lipkin, Phys. Lett. B214 (1988) 429}
$[{\HJL}]$
to express the
total quark contribution to the proton spin in terms of $E$ and $N$
which are directly measured and $\Delta s$ whose value satisfies the
inequality (12).
$$ \Delta q \equiv \Delta u + \Delta d + \Delta s =
{18\over 5}\cdot E - {3\over 5}\cdot N
+ {3\over 5}\cdot \Delta s.
\eqno (13)$$
Then
$$ \Delta q (\epsilon = 0) \leq \Delta q \leq
\Delta q (\epsilon = 0) - {3\over 5}\cdot \Delta s. \eqno (14a) $$
Substituting the results (10) gives
$$
0.12 \pm 0.17 \leq \Delta q \leq 0.23 \pm 0.17
\eqno (14b) $$
Thus the correction to $\Delta q$ is still less than one standard
deviation even in the limit of large SU(3) breaking which reduces
$\Delta s$ to zero.

Further insight into the roles of valence and sea quarks in weak decays
is illustrated by comparing the information obtained from these decays
with that obtained from deep inelastic scattering. The EMC result (7)
weighs quark and antiquark contributions equally and therefore also
weighs valence and sea contributions equally.
In deep inelastic scattering of a polarized $W^-$ beam on a nucleon
target where one observes inclusively all final states with a given
strangeness, the contributions from both the quarks and antiquarks in
the nucleon are equally weighted. The results for $W^-p$ inclusive
scattering into all nonstrange final states therefore give the quantity
$\Delta u(p) - \Delta d(p)$. Similarly $W^-n$ inclusive scattering into
all strange final states gives $\Delta u(n) - \Delta s(n)$.

We now examine whether the weak decays give the same information
and in particular
whether the quark and antiquark transitions or the
valence and sea transitions at the quark level are equally weighted.
We can compare the relative weighting of valence and sea contributions
obtained from weak decays and from deep inelastic $W^-$ scattering by
noting that the
neutron and $\Sigma^-$ decays are related respectively by time reversal
to exclusive neutron and $\Sigma^-$ production in $W^- N$ scattering,
$$ W^- + p \rightarrow n; \;\;\;\;
W^- + n \rightarrow \Sigma^-  \eqno (15)    $$

A sea transition changes the electric charge of the sea.
In inclusive scattering this causes no problem, all final states are
considered. But in the exclusive transitions (15) the sea contribution
can be reduced by a wave-function overlap factor between
initial and final states with seas having different charges.

If the sea is flavor symmetric, the sea contributions to
$\Delta u(p) - \Delta d(p)$ or $\Delta u(n) - \Delta s(n)$ exactly
cancel. Thus the relative weighting of sea and valence contributions
is irrelevant to the determination of these differences and the same
results are expected in inclusive transitions and in the exclusive
reactions (15).

If the sea is not isospin symmetric, there is a
sea contribution to $\Delta u(p) - \Delta d(p)$. But isospin invariance
then requires the proton wave function to have a component where
the electric charge of the valence quarks is zero and the proton charge
is carried by the sea and predicts the same result for
$\Delta u(p) - \Delta d(p)$ from neutron decay and deep inelastic
scattering. This is confirmed by our result (2).

Similarly, if the sea is not SU(3) symmetric SU(3) symmetry
would requires the $\Sigma^-$ wave function to have a component where
the strangeness of the valence quarks is zero and the strangeness
is carried by the sea. This is not expected; there
should be no sea contribution to $\Sigma^-$ decay. Thus we see again
that $\Delta u(n) - \Delta s(n)$ should be given by
the result (4) in which only valence quarks contribute, This argument
for the results (3-11) is more general, since
it does not assume that the seas in the neutron and $\Sigma^-$ are the
same. The results thus apply even if the sea is not isospin symmetric
and the isospin asymmetries in the neutron and $\Sigma^-$ are
different.
Note that eq. (5) relates only $\Delta u(n)$ and $\Delta s(n)$
which are equal by isospin to $\Delta d(p)$ and $\Delta s(p)$ but makes
no assumption about $\Delta u(p)$ and $\Delta d(n)$.

The other hyperon decays can be treated with SU(3) symmetry breaking
by analogy with eqs. (3-5) if we assume that the sea is isoscalar,
$$ \Xi \equiv {{G_A}\over{G_V}}(\Xi^- \rightarrow \Lambda) =
{1\over 3}[ \Delta u (p)+ \Delta d (p) -2 (1+\epsilon)\cdot \Delta s (p)]
\eqno(16a)
$$
$$ \Lambda \equiv
{{G_A}\over{G_V}} (\Lambda \rightarrow p) =
{1\over 3}[2 \Delta u (p)- \Delta d (p) - (1+\epsilon)\cdot \Delta s (p)]
\eqno (16b) $$
where we have again introduced the shortened notation $\Xi$ and
$\Lambda$ .
The experimental values can be compared with the SU(3) predictions
using the $F$ and $D$ values obtained from eqs. (6).
$$ 0.25 \pm 0.05 = \Xi = F-{D\over 3} = 0.193 \pm 0.009 \eqno(17a)
$$
$$ 0.718 \pm 0.015 = \Lambda = F+{D\over 3} = 0.729 \pm 0.009
\eqno (17b) $$

We also obtain two additional independent values for the linear
combination (6a) which transforms like the isoscalar component of an
octet,
$$ {{3F - D} \over{\sqrt 3}} =
2 \sqrt 3 \cdot   \Lambda  -
 \sqrt 3  \cdot
N =
0.310 \pm 0.05
$$
$$ =  \sqrt 3 \cdot
\Xi =
 0.43 \pm 0.09
\eqno (18) $$

The experimental values are in agreement with SU(3), and again the
effect of SU(3) breaking on information obtained about proton spin
structure is to replace $\Delta s$ by $(1+\epsilon) \Delta s$.
These results, even with SU(3) symmetry breaking, still satisfy
the SU(3)
``equal spacing rule" \REF{\FRANKLIN}{Z. Dziembowski and J. Franklin,
J. Phys. G17 (1991) 213} [{\FRANKLIN}].
$$
(1/3)[ \Delta u (p)- 2\Delta d (p) - (1+\epsilon)\cdot \Delta s (p)]
= $$
$$ = N- \Lambda = \Lambda  - \Xi = \Xi - \Sigma =
{1\over 3}\cdot (N- \Sigma)
\eqno (19a) $$
$$ 0.439 \pm 0.015 =  0.47 \pm 0.05 = 0.60 \pm 0.05 = 0.532 \pm 0.006
\eqno (19b) $$

These equations show that breaking SU(3) only by suppressing the strange
quark contribution in an isoscalar sea leaves the SU(3) relations between
the different hyperon decays intact, and affects only the relation
between
the decay rates and the spin structure of the baryons by introducing the
factor $(1+\epsilon)$ in the coefficient of $\Delta s$ everywhere.

However, the experimental errors are so large in the $\Xi$ and $\Lambda$
decays that using them together
with the other data to determine the D and F parameters and the isoscalar
component (6a) provides a negligible improvement. This is seen most
easily in the equal-spacing parameter (19) where the error in the value
obtained by using only the neutron and $\Sigma^-$ decays is much smaller
than in values obtained by using other decays. Thus even in unbroken
SU(3) it is simpler not to use the F and D parametrization, deal directly
with the data and use the experimental fact that the error in
$ n\rightarrow p$ is much smaller than in all other decays.
Better experimental values, especially for
$ {{G_A}\over{G_V}}(\Xi^- \rightarrow \Lambda)$ would provide a
significant test of SU(3) breaking, as well as providing more insight
into the $n-\Lambda-\Sigma^-$ paradox
$[{\PNAGOYA}]$.

We also note that our results obtained by using only the
$ \Sigma^-\rightarrow n $ data do not assume an isoscalar sea, while
the results (19) are valid only for an isoscalar sea. Thus if
there is evidence that the sea is not isoscalar it is also safer to
use only $ \Sigma^-\rightarrow n $ data.

Stimulating and clarifying discussions with Marek Karliner and Jechiel
Lichtenstadt are gratefully acknowledged.
This research was partially supported
by the Basic Research Foundation administered by the Israel Academy of
Sciences and Humanities and by grant No. 90-00342 from the United
States-Israel Binational Science Foundation (BSF), Jerusalem, Israel.

\refout

\end